\documentclass[fleqn,10pt]{wlscirep}
\usepackage[utf8]{inputenc}
\usepackage[T1]{fontenc}
\usepackage{amsmath}
\usepackage{amsfonts}
\usepackage{latexsym}
\usepackage{amssymb}
\usepackage{geometry}
\usepackage{graphicx}
\newcommand{\lcg}{\mbox{LaCrGe$_3$}}
\newcommand{\ug}{UGe$_2$}

\newcommand{\ui}{UIr}
\newcommand{\zr}{ZrZn$_2$}
\newcommand{\g}[1]{{\bf #1}}

\newcommand{\dagga}{{\phantom{\dagger}}}

\title{Mechanism for transitions between ferromagnetic and antiferromagnetic orders in $d$-electron metallic magnets}

\author[1,*]{Marcin M. Wysoki\'nski} 
\affil[1]{International Research Centre MagTop, Institute of Physics, Polish Academy
of Sciences, Aleja Lotnik\'ow 32/46, PL-02668 Warsaw, Poland}
 
\affil[*]{wysokinski@magtop.ifpan.edu.pl}

\begin{abstract}
We propose mechanism for  pressure-induced transitions between ferromagnetic and antiferromagnetic phases that relies on a competition between characteristic energy scales ubiquitous among $d$-electron metallic magnetic compounds. Principles behind the mechanism are demonstrated on the example of the minimal two-orbital $p$-$d$ lattice model. 
We suggest that \lcg, where pressure-induced ferromagnetic-to-antiferromagnetic phase transition has been recently observed, is a promising candidate to realize discussed mechanism. 
\end{abstract}
\begin{document}

\flushbottom
\maketitle
 
\thispagestyle{empty}

\section*{Introduction}

Several experiments have provided evidences for magnetic groundstate switching between ferromagnetic (FM) and antiferromagnetic (AFM) orderings driven by pressure or chemical substitution in $d$ and $f$ electron compounds \cite{Sven2018,Taufour2016,Taufour2017,Jeffries2016,Aoki2017,geibel2013}.
Nevertheless, there exist only few theoretical studies aiming on describing these magnetic transitions.

Recently, general expectations concerning order of transitions between FM and AFM phases have been provided within extended Landau theory \cite{Belitz2017}. 
In turn, the generic microscopic mechanisms triggering such transitions, except   material specific ones \cite{Wysokinski2018R,Leonid2019},  are mostly unidentified. The effective action approach incorporating corrections due to quantum fluctuations \cite{Belitz1997,Green2018} constitutes a notable exception. Namely, it has been shown that in the vicinity of the ferromagnetic quantum critical point quantum fluctuations favor reconstruction of the phase diagram and the first-order transition at zero temperature or emergence of the spatially modulated magnetic phase is predicted  \cite{Belitz1997,Chubukov2004,Conduit2009,Karahasanovic2012,Green2018}.     
Experimental observations are consistent with both scenarios\cite{Brando2016,Pfleiderer2002,Uhlarz2004,Taufour2010,Kotegawa2011,geibel2013,Taufour2016}.

\begin{figure}[t]
  \begin{center}   
    \includegraphics[width= 0.45  \textwidth]{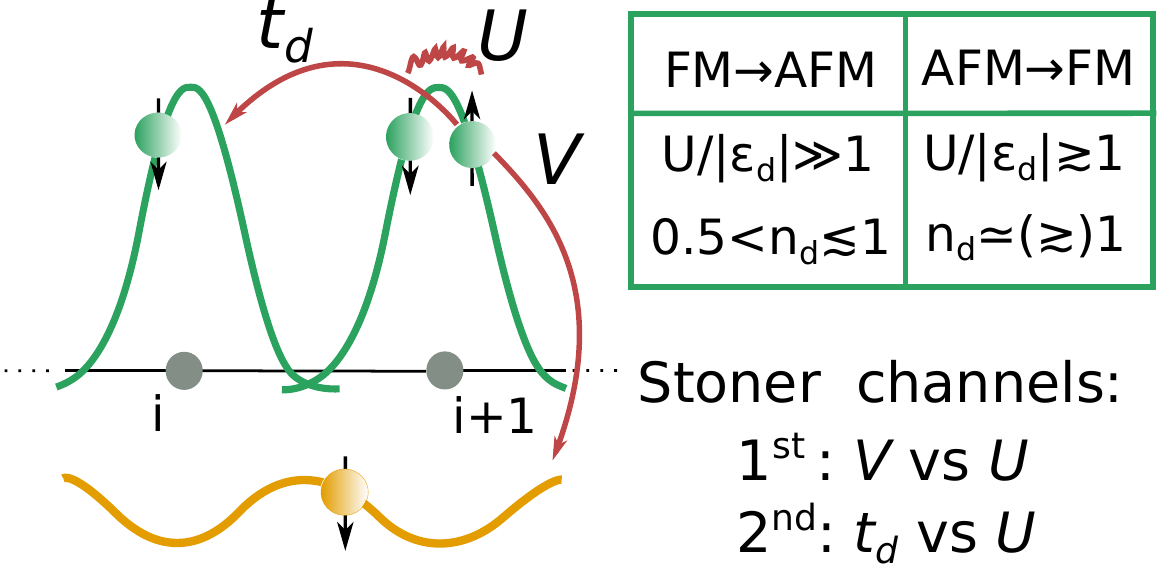}
  \vspace{-0.2  cm}
  \end{center} 
 \caption{Schematic orbital structure of the Hamiltonian \eqref{H} providing strongly localized $d$-orbital states (green) and delocalized $p$-ones (yellow). Competition between intra-$d$-orbital interaction  ($U$) and, either inter-orbital hybridization ($V$) or direct $d$-orbital hopping ($t_d$), accounts for the first and the second {\it  Stoner channels}, respectively. The table summarizes general necessary (not sufficient) conditions to be fulfilled by $d$-orbital at ambient pressure state to allow for a subsequent magnetic groundstate switch (cf. main text).}
\label{fig0} 
\end{figure}

In this work, we propose a different mechanism for transitions between itinerant FM and AFM orders that equally applies to nominal ferromagnets as well as antiferromagnets.  It relies on a common for  metallic magnetic compounds low-energy electronic structure deriving from  correlated  orbital states hybridizing to rather uncorrelated  ones (cf. Fig. \ref{fig0}).  Such situation is indicated by {\it ab-initio} calculations 
for several itinerant magnets, including \lcg, \zr, CrAs, \ug\ and \ui\ \cite{Jarlborg1981,Jarlborg2001,Singh2002,Bie2007,Nguyen2018, Autieri2017,Noce2017,Autieri2018,Onuki2010,Shick2001,Samsel2011}. We propose that the Fermi-liquid description of such general orbital structure can lead, exclusively in $d$-electron magnets, to pressure-induced transitions between FM and AFM orders.

In order to demonstrate the advocated, as we call it, {\it two-channel Stoner mechanism} we analyze simple $p$-$d$ model that account for a mentioned energy scales in a minimal manner. We show that both FM/AFM and AFM/FM transitions can be realized by the model in a specific regime of parameters related to the character of $d$-orbital such as: degree of correlations $U$, $d$-level position $\varepsilon_d$ and $d$-orbital filling $n_d$ (cf. table of Fig. \ref{fig0}).
It is very promising that the parametrization roughly agreeing with general energy scales and oxidation states of \lcg\ qualifies observed in this material pressure-induced FM to AFM transition \cite{Taufour2016,Taufour2017} as a potential manifestation of the {\it two-channel Stoner} physics. 

\section*{Model}
The  $p$-$d$ lattice Hamiltonian, 
\begin{equation}
 \mathcal{H} = \sum_{\g k\sigma} \epsilon^d_{\g k} \hat d_{\g k\sigma}^\dagger\hat d^\dagga_{\g k\sigma}+U\sum_{\g i}\hat n_{\g i\uparrow}^d \hat  n_{\g i\downarrow}^d +\sum_{\g k \sigma} \epsilon^c_{\g k} \hat c_{\g k\sigma}^\dagger\hat c^\dagga_{\g k\sigma}+\sum_{\g k \sigma} \big(V_{\g k} \hat d_{\g k\sigma}^\dagger\hat c^\dagga_{\g k\sigma}+{\rm H.c.}\big),
\label{H}
\end{equation}
describes in a minimal manner mixing between correlated $d$ (operators $\hat d$) and uncorrelated $p$ (operators $\hat c$) orbital states, scenario ubiquitously present in metallic $d$-electron magnets.
The stronger degree of electronic correlations among $d$ than $p$  states is attributed to the more localized nature of the former ones   and is accounted for by the presence of the onsite Coulomb repulsion with the amplitude $U$ in the $\hat d$-operator subspace, whereas lack thereof in  the $\hat c$-operator subspace. In \eqref{H} $\sigma\in\{\uparrow,\downarrow\}$ is a spin index, {\g i} is a position vector of an underlying lattice of  correlated  orbitals, $\g k$ is a momentum vector in the first Brillouin zone and $\hat n^d_{\g i\sigma}\equiv \hat d^\dagger_{\g i\sigma} \hat d^\dagga_{\g i\sigma}$.  

The $p$-$d$ Hamiltonian in a present work serves for a demonstration purposes of the principles behind {\it two-channel Stoner mechanism} rather than modeling of any realistic electronic structure. Therefore,
without referring to particular orbital structure of $d$ and $p$ states, we assume that intra-orbital kinetic energies $\mathcal{E}_{\g k}$ and $\epsilon_{\g k}$, of the correlated and the uncorrelated subsystems respectively, have a dispersion proportional to $\xi_{\g k}\equiv-2(\cos{k_x}+\cos{k_y})$, i.e., $\epsilon^{\alpha\in\{c,d\}}_{\g k}=\epsilon_\alpha+t_\alpha\xi_{\g k}$ where  we set $\epsilon_c=0$. 
Moreover, inter-orbital hybridization is assumed in a momentum independent form ($V_{\g k}=V$) and the $p$-$p$ hopping is taken as the energy unit, $|t_c|=1$. Finally, we adopt here the condition $t_d/t_c<0$, that, unless specific geometrical reasons enter the problem, is typical case for direct bondings between orbital states differing with an angular momenta by 1. It is worth mentioning that mixed $p$-$d$ quasiparticle density of states is dominated by features originating from hybridization between orbitals rather then these related to the underlying lattice. In that manner, even perfect nesting for each of the orbital subsystem in quasiparticle spectrum is inherited only at integer values of total filling. Therefore, made in our work choice of total-filling away from integer,  for above assumption  leads to hole- or electron-like Fermi-surfaces.  

For the forthcoming discussion it is important to note that the $p$-$d$ model for $t_d=0$ reduces to the Anderson lattice model which describes mixing of localized $f$-states with delocalized conduction states and is frequently invoked for the description of heavy fermion systems. On the other hand for $V=0$, orbital states in $p$-$d$ model are not mixing and thus system of $d$-orbital states is described by the Hubbard model, which constitutes the simplest description of correlated electron systems, where complexity of real material is reduced just to the competition between the on-site repulsion U and hopping $t_d$.

\section*{Two-channel Stoner mechanism}

Superficially one could expect that magnetism in itinerant magnets occurs by means of the usual Stoner mechanism due to a competition between the potential energy of correlated electrons  and the total kinetic energy. 
However, such an approach  overlooks in $d$-electron magnets such as \lcg, \zr\ or CrAs  displaying mixed correlated/uncorrelated electronic structure \cite{Jarlborg1981,Jarlborg2001,Singh2002,Bie2007,Nguyen2018, Autieri2017,Noce2017,Autieri2018,Uhlarz2004,Taufour2016} existence of two, qualitatively different {\it Stoner channels}.

 First {\it Stoner channel} ($1^{st}$SC)  accounts for a competition between strong interactions among $f$ or $d$ states and their hybridization  to the weakly correlated ligand states ($U$ vs $V$, cf. Fig. \ref{fig0}). In turn, the second {\it Stoner channel} ($2^{nd}$SC)  refers to the competition between the same interactions and the kinetic energy due to a direct metallic bonding between correlated orbital states ($U$ vs $t_d$, cf. Fig. \ref{fig0}). $2^{nd}$SC is usually absent in $f$-electron metallic magnets due to a negligible intra-$f$-orbital hopping  \cite{Shick2001,Onuki2010,Samsel2011,Wysokinski2014R,Wysokinski2015R} ($t_d\simeq0$). Contrary, it is clearly present in $d$-electron magnets such as \lcg, \zr\ and CrAs \cite{Jarlborg1981,Jarlborg2001,Singh2002,Bie2007,Nguyen2018,Autieri2017,Autieri2018,Noce2017}.
  
Intuition for the character of each channel separately can be gained by recalling  magnetic properties of mentioned earlier, two well-established models: two-band Anderson lattice model \cite{Doradzinski1997,Nolting2000,Bonca2002,Bonca2003,Byczuk2008,Kubo2013,Wysokinski2014R,Wysokinski2015R,Kubo2015,Tremblay2015,Aulbach2015,Abram2016,Wysokinski2018R} (Eq. \eqref{H} with $t_d=0$) with only $1^{st}$SC present, and single-band Hubbard model \cite{Kotliar1986,Hirsch1985,Vollhardt1999,Lichtenstein2000,bunemann2016,Tocchio2016} (Eq. \eqref{H} with $V=0$) accounting solely for $2^{nd}$SC. Both models for moderate interactions (limit allowing to disregard local moment magnetism related either to Kondo or Mott physics)  favor FM when total filling $n_t$ is away from  integer. On the other hand, $n_t$ close to integer due to combined effect of nesting and stronger correlation effects yield stable AFM.      

The situation in a $p$-$d$ system, with both channels active, is more subtle. In such case the correlated band-filling $n_d$ is not fixed, in contrast to $n_t$, and can notably change in a response to a modification of parameters associated with an applied pressure. Consequently, in response to a modification of $n_d$, the character of $2^{nd}$SC possibly can change in the decisive, for a favored magnetic ordering, manner.
In a following, we thoroughly explore this idea with the quantitative analysis of $\mathcal{H}$.

   \begin{figure}[ t]
  \begin{center}   
      \includegraphics[width=   \textwidth ]{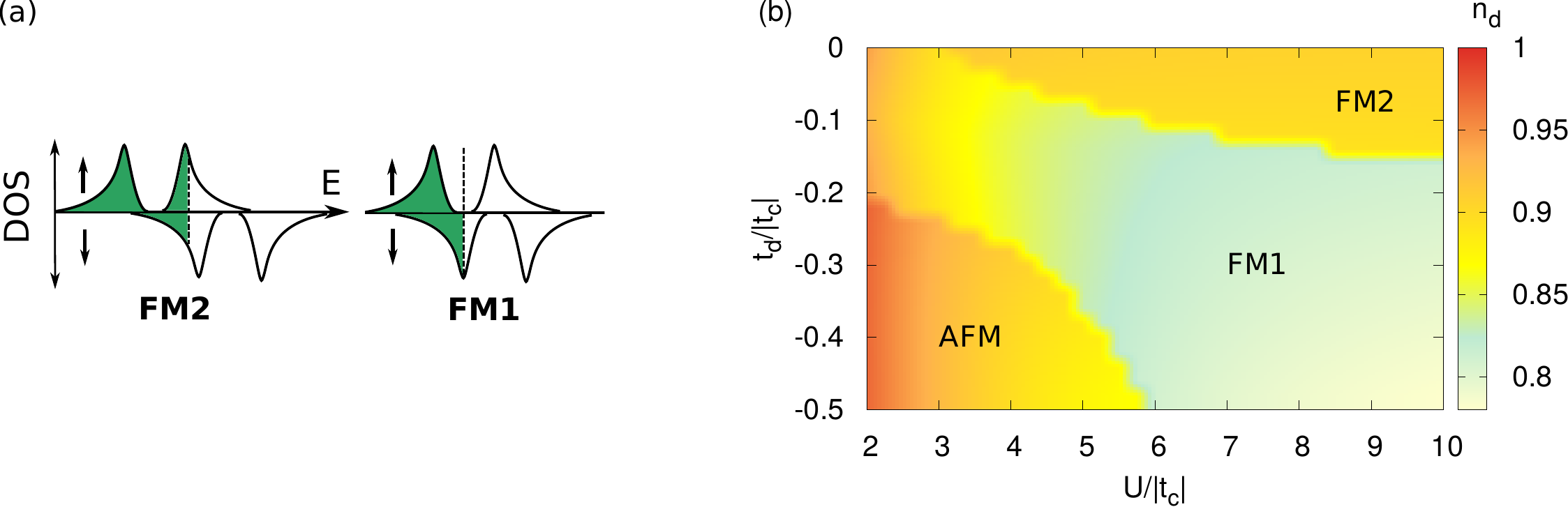}   
    \end{center} 
\vspace{-0.5cm} 
\caption{(a) Schematic picture of the spin-resolved density of states corresponding to two ferromagnetic phases, FM2 and FM1 that differ with Fermi surface topology due to spin selective gap in the former phase. (b)
Magnetic phase diagram on the  interaction - $d$-hopping, $U$-$t_d$ plane for $\varepsilon_d=-2$, $V=-0.9$ and $n_t=1.6$. Color scale       refers to the $d$-orbital filling, $n_d$.
} 
\label{fig} 
\end{figure}  

\section*{Results}
In this section we analyze magnetic properties of the $p$-$d$ lattice model, which constitutes a minimal set-up accommodating  two {\it Stoner channels}. Precisely, Hamiltonian \eqref{H}  is treated within modified renormalized mean-field theory \cite{Wysokinski2014,Wysokinski2014R,Abram2017} (see also Methods section) for a stability of three phases conveniently characterized with an average, spin and space dependent, density $\langle \hat n_{\g i \sigma}\rangle$: PM Fermi-liquid with $n^{PM}_{\g i \sigma}\!=\!n_t/2$; uniformly polarized FM  phases with $n^{FM}_{\g i \sigma}\!=\!(n_t\!+\!\sigma m_{FM})/2$; and spatially modulated AFM phase with $n^{AFM}_{\g i \sigma}\!=\! (n_t\!+\!\sigma m_{AF} {\rm e}^{i \g Q \g R_{\g i}})/2$ with usual Ne\'el vector ${\g Q\!=\! (\pi,\pi)}$ and $m_{FM}$ and $m_{AF}$ corresponding to  uniform and staggered magnetization, respectively. Presence of the indirect hybridization gap in the spectrum allows for two FM phases, FM1 and FM2 differing with a Fermi surface topology schematically drawn in  Fig. \ref{fig}a \cite{Wysokinski2014R,Kubo2013}.
We note that such a distinction provides a natural rationalization to experimentally established in several metallic ferromagnets (\lcg, \zr, \ug) two different FM1 and FM2 phases in the $p$-$T$-$h$ diagrams \cite{Pfleiderer2002,Uhlarz2004,Kimura2004,Taufour2017}.

In  Fig. \ref{fig}b we present magnetic phase diagram on the interaction - $d$-hopping plane with color scale denoting $d$-orbital valence. 
Parametrization is chosen in order to demonstrate advocated scenario of two Stoner channels, and tunable character of the $2^{nd}$SC.
Using the notions from the 
conceptual considerations in previous section, we may describe obtained magnetic phase diagram in a following manner. Total filling away from integer value fixes $1^{st}SC$ to favor FM. Character of $2^{nd}$SC is in turn controlled by $t_d$ and $n_d$. In the case of non-active $2^{nd}$SC ($t_d=0$) in Fig. \ref{fig}b there is realized FM state supported by $1^{st}SC$, even when $d$-orbital filling is close to $n_d=1$. 
On the other hand, for substantially large $|t_d|$ and when $n_d$ approaches half-filling in Fig. \ref{fig}b AFM phase becomes stable. This is because in this limit, not only $2^{nd}$SC favors AFM order but also  the tendency toward FM in $1^{st}SC$ is overcome. 

Next, we search for the parameter space of the model in a presence of active $2^{nd}$SC that could allow for the magnetic phase transitions possibly associated with applied pressure.   Due to the essentially more localized nature of $d$ than $p$ orbital states hopping integrals in the following discussion are parametrized by $|t_d|=|t_c|/8$. Effects of an applied pressure in  $\mathcal{H}$ can be reasonably captured by assuming identical increase of all {\it kinetic} amplitudes, $t_c$, $t_d$ and $V$, provided fixed $U$ and $\varepsilon_d$. In the case of $|t_c|=1$ playing a role of an energy unit, above convention for emulating applied pressure follows  decreasing ratios of $U/|t_c|$ and $|\varepsilon_d|/|t_c|$, with all kinetic amplitudes unchanged.

Intriguingly, decrease of each ratio separately can influence $n_d$, and thus $2^{nd}$SC, in an opposite manner. Namely, it is clear that change of $\varepsilon_d/|t_c|<0$  under pressure
towards 0 modifies balance between orbital-resolved fillings providing lower $n_d$. On the other hand,  weakening correlations (decreasing $U/|t_c|$), due to a reduction of the energy penalty for double occupancies, allows for its increase. The later relationship is clearly visible in Fig. \ref{fig}b where decrease of $U/|t_c|$, except FM2/FM1 transition, is accompanied with an increase of $d$-filling.

Shift of $\varepsilon_d$ towards 0 due to an applied pressure can be of negligible importance in comparison to decreasing interactions if 
$U/|\varepsilon_d|\gg1$. Given that the ambient pressure state is characterized by $n_d\lesssim 1$, weakening correlations allow for a larger $n_d\simeq1$. In principle, such situation can lead to the change of the ordering favored by the $2^{nd}$SC from FM to AFM. In principle this change could be decisive for the realized state and thus pressure-driven FM/AFM transition in such scenario could be expected.

 \begin{figure}[t]
  \begin{center}   
      \includegraphics[width =   \textwidth,trim={0cm 0cm 0cm 0.0cm},clip]{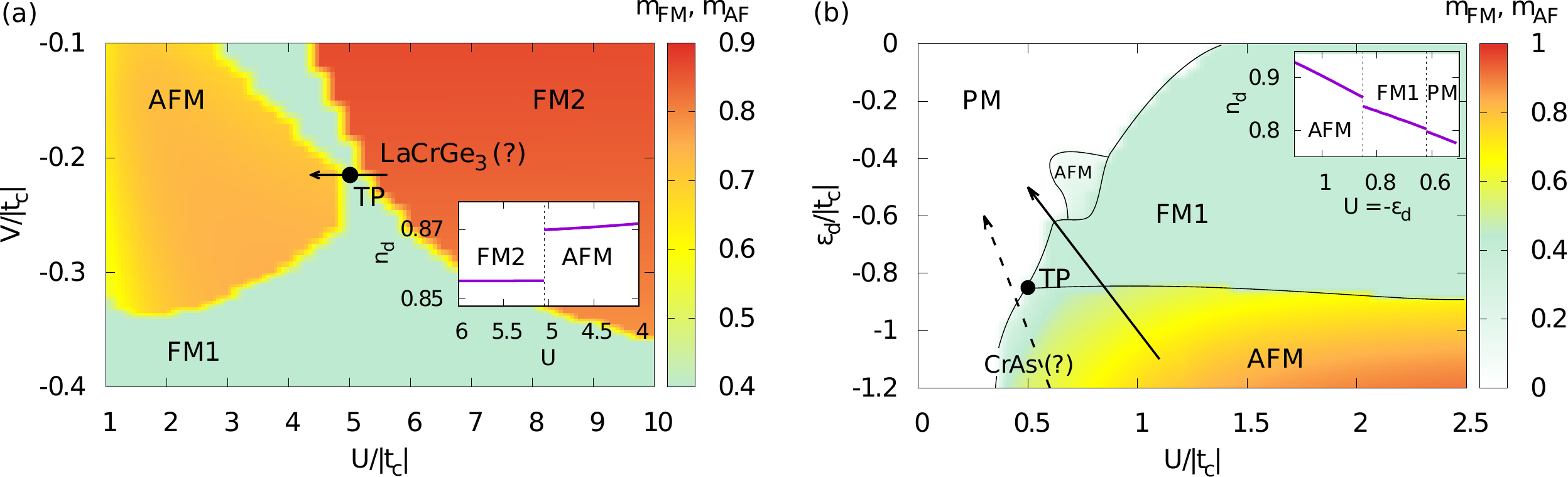}  
  \end{center} 
\vspace{-0.5cm} \caption{Magnetic phase diagrams (a) on the interaction - hybridization, $U$-$V$ plane for $\varepsilon_d=-0.95$ and (b) on the interaction - $d$-level, $U$-$\varepsilon_d$ plane for  $V=-0.25$ (in both cases $n_t=1.6$). Color scale denotes total ordered magnetic moment in each phase. Arrows mark a possible directions  associated with an applied pressure (cf. main text).
Insets show changes of $d$-orbital valence along solid-line arrows  
encompassing, (a) FM/AFM and (b) AFM/FM, transitions.
} 
\label{fig2} 
\end{figure}  

In Fig. \ref{fig2}a we present magnetic phase diagram on the $U$-$V$ plane for the relatively shallow $d$-level, $\varepsilon_d=-0.95$. Indeed, we have obtained FM to AFM transitions driven by decreasing $U/|t_c|$.  As we checked, these transitions are always associated with increase of $n_d$ towards half-filling in AFM phase, in accord with the proposed two channel Stoner mechanism,  suggesting in such case strengthening tendency  in  $2^{nd}$SC to favor AFM.
In Fig. \ref{fig2}a we have marked with the arrow exemplary direction ($V=-0.21$) mirroring application of pressure encompassing triple point (TP) between FM1, FM2 and AFM phases for $U\simeq5$, assuring  the condition $U/|\epsilon_d|\gg1$ is fulfilled. 
In the inset of Fig. \ref{fig2}a we explicitly show  jump  of $d$-orbital filling toward the half-filling  at FM/AFM transition along marked arrow.

The pressure-driven FM/AFM transition predicted by the proposed  mechanism is rather fragile. This is because
condensation energy of AFM state with respect to FM1 phase is small in the whole AFM stability region.  
In this manner, even a small Zeeman splitting  due to a magnetic field  would quickly decrease parameter-space area with stable AFM phase in favor of FM1 (but not FM2).
Similar effect, but of different origin, has further depletion of the $d$-orbital filling towards $n_d=0.5$ by lifting  $d$-level from $\varepsilon_d=-0.95$ (used in Fig. \ref{fig2}a) to $\varepsilon_d=-0.8$. It yields stable FM1 phase instead of AFM  along the same direction ($V=-0.21$). It happens because in the diluted $d$-filling regime, irrespectively of the $U$ value, the average number of double occupancies is negligible $\langle\hat n^d_\uparrow\hat n^d_\downarrow\rangle\simeq0$. Consequently, decrease of the energy penalty, originating from $U\langle\hat n^d_\uparrow\hat n^d_\downarrow\rangle$ contribution to the free energy, is not efficient enough to allow for an increase of $n_d$ triggering transition to AFM.

On the other hand, in a system which obeys $U/|\varepsilon_d|\simeq1$, applied pressure should modify both $|\varepsilon_d|/|t_c|$ and $U/|t_c|$ ratios on an equal footing such that $U/|\varepsilon_d|$ is fixed. In such a scenario an applied pressure is expected to decrease $d$-filling, due to likely leading effect of upshift of $\varepsilon_d$. Given that the ambient pressure state is characterized by $n_d\gtrsim1$ (e.g. due to deep $d$-level) and favors AFM, the decrease of $d$-filling  strengthens tendency toward FM in $2^{nd}$SC. It indicates a possibility of a pressure-driven AFM/FM transition.

In Fig. \ref{fig2}b we present a magnetic phase diagram on the $U$-$\varepsilon_d$ plane for $V=-0.25$ which demonstrates that achieving such transition is possible.
As we checked, all AFM/FM transitions along directions with fixed  ratio $U/|\varepsilon_d|$ are  associated with decrease  of $n_d$ away from half-filling, in accord with the scenario in which character of $2^{nd}$SC changes together with $d$-orbital filling.
The solid arrow in Fig. \ref{fig2}b, drawn for $U/|\varepsilon_d|=1$ determining a possible direction associated with an applied pressure, crosses AFM/FM transition. In the inset of Fig. \ref{fig2}b we explicitly show the evolution of the $d$-orbital filling along this direction. 
 
 In the table in Fig. \ref{fig0} we have summarized, established by the prior analysis of $\mathcal H$, conditions to be fulfilled by $d$-orbital at ambient pressure indicating possibility for subsequent pressure-induced FM/AFM or AFM/FM transitions supported by the {\it two-channel Stoner} mechanism. 

  \section*{Candidate compounds} 
The most promising candidate to realize magnetic phase transition due to {\it two-channel Stoner} physics is \lcg.  
Face-sharing octahedra of Ge atoms encompassing Cr in \lcg\ imply that correlated $3d$-states not only hybridize to $p$-ones of surrounding ligands  but also bond between each other \cite{Bie2007}, justifying a general $p$-$d$ orbital structure of the model \eqref{H}. In turn,  parametrization of the model can be extracted from the electronic properties of \lcg\: ({\it i}) shallow $d$-level position \cite{Bie2007,Nguyen2018} and $3d$ nature of Cr valence states indicate $U/|\varepsilon_d|\gg1$; ({\it ii}) general agreement with (La$^{3+}$)(Cr$^{3+}$)(Ge$^{2-}$)$_3$ oxidation state \cite{Bie2007} pinpoints $n_d<1$; ({\it iii}) bonding between $3d$-states twice smaller than their hybridization to $p$-states \cite{Bie2007} suggests $2|t_d|\simeq|V|$.

All these features fulfill general conditions (cf. first column of table in Fig. \ref{fig0}) for a pressure-induced FM/AFM transition (cf. arrow in Fig. \ref{fig2}a) qualifying recent observation of magnetic groundstate switching in \lcg\ \cite{Taufour2016,Taufour2017}  as a possible manifestation of the {\it two-channel Stoner} physics. The incorporation of the detailed electronic structure of this material is clearly needed to further support proposed interpretation. Nevertheless, experimental discrimination between large ($Q\sim \pi$) or small ($Q\sim0$) ordering vector of the AFM state in \lcg\ is highly desirable. Large vector would be consistent with {\it two-channel Stoner} physics whereas small one with the mechanism based on the quantum fluctuation arguments \cite{Conduit2009,Karahasanovic2012,Green2018}.

We also note that the realization of TP between FM1, FM2 and AFM phases (cf. Fig. \ref{fig2}a) \cite{notkas}, 
although constitutes very appealing rationalization of a similar FM1-FM2-AFM meeting point observed in \lcg\ at 2 K \cite{Taufour2017}, is an effect of a fine-tuning rather than a robust feature of the model.
Nevertheless, field-driven increase of the stability region of FM1 phase at the expense of the AFM phase already is a generic property. It is related to the weak condensation energy of AFM phase with respect to FM1 (but not to FM2) over wide range of $U$ values. In that manner, prediction of our simple model agrees with observations in  \lcg\ where stability region of FM1 emerges out of AFM phase with application of pressure.    

Second promising candidate material to realize {\it two-channel Stoner} physics is itinerant $d$-electron  antiferromagnet CrAs.
Band structure calculations for this material suggest:
({\it i})  predominant role of mixed $p$-$d$ states near the Fermi level \cite{Autieri2017,Noce2017,Autieri2018}; ({\it ii})  proximity between FM and AFM states \cite{Autieri2017}; ({\it iii}) weak degree of correlations  due to a very wide $p$-band (large $|t_c|$) \cite{Autieri2017,Noce2017,Autieri2018}; ({\it iv}) consistency with (Cr$^0$)(As$^0$)  oxidation state, implying occupation of $3d$-orbital larger but close to half-filling \cite{Noce2017}.

First two properties ({\it i-ii}) of CrAs indicate a possible importance of the {\it two-channel Stoner} mechanism to describe properties of this material. The third one ({\it iii}) suggests that due to the small ratio $U/|t_c|$, both $U$ and $\epsilon_d$ change under pressure on an equal footing (cf. Fig. \ref{fig2}b). Finally, the forth one ({\it iv}) indicates rather deep $d$-level such that $U/|\varepsilon_d|<1$. Consequently, the direction possibly associated with CrAs under pressure is drawn in Fig. \ref{fig2}b with labeled dashed-arrow for $U/|\varepsilon_d|=0.5$.  The proximity to AFM/FM transition, present in Fig. \ref{fig2}b already for $U/|\varepsilon_d|\simeq0.6$,  can justify why {\it ab-initio} calculations,  that are known to slightly overestimate tendency toward FM, support stable FM in CrAs under pressure \cite{Autieri2017}.

Finally, we note that the similar minimal model proposed for itinernat ferromagnet UGe$_2$ \cite{Wysokinski2014R,Wysokinski2015R,Abram2016} not only have rationalized compound's magnetic properties but it inspired proposal on the feasible origin of the triplet superconductivity \cite{Ewa2018,Fidrysiak2019}  based on the Hund's rule induced pairing \cite{Spalek2001,Zegrodnik2012,Zegrodnik2013,Zegrodnik2014}.
Therefore, it is tempting to ask whether our model  can also shed some light on the possible origin of superconductivity in CrAs at the border of the magnetic/non-magnetic phase \cite{Wei2014}. 
In fact, inspired by the recent experimental results in the similar class of Cr-based compounds \cite{luo}, we could speculate that the potential vicinity of the ferromagnetic instability in CrAs and thus ferromagnetic fluctuations can be a candidate pairing glue.

\section*{Summary} 
In the present work  we have proposed  mechanism for competing AFM and FM orders in metallic systems displaying mixed correlated $d$-orbital/uncorrelated ligand-orbital electronic structure in the vicinity of the Fermi level. We have analyzed magnetic instabilities of the Fermi liquid realized by a minimal two-band model in order to demonstrate properties of the mechanism.
We have formulated general conditions (cf. table of Fig. \ref{fig0}) that  relate energy scales connected with the character of  $d$-orbital at ambient pressure with potentially realized FM/AFM or AFM/FM transitions in response to applied pressure. We found, that the character of $d$-orbitals in \lcg\ is consistent with the proposed here parametrization of the model supporting pressure induced FM/AFM transition. It is very promising result in terms of potential interpretation of recent observations of such transition in this material \cite{Taufour2016,Taufour2017}.

\section*{Methods}

The low-energy  properties of the $p$-$d$ model in the $T\rightarrow0$ limit, i.e. the Fermi-liquid state and its magnetic instabilities, can be efficiently analyzed within the Gutzwiller approximation combined with the optimization of the Slater determinant scheme \cite{Wysokinski2014,Wysokinski2014R,Abram2017}. The technique belongs to the renormalized mean-field theory class of approaches \cite{Shiba,Gebhard2007}. 
Formally, the method accounts for the variational optimization of the average number of double occupancies $\langle\hat n^d_\uparrow\hat n^d_\downarrow\rangle$  with the Gutzwiller correlator. In practice, it reduces to the construction of the effective single-particle mean-field Hamiltonian for each considered phase with renormalized with Gutzwiller factor $q_\sigma$ certain characteristics such as hybridization and $d$-$d$ hopping \cite{notka}.

Renormalized mean-field Hamiltonian for FM phase reads
\begin{equation}
 H_{FM}=\sum_{\g k,\sigma} \Psi^\dagger
 \begin{pmatrix}
  \epsilon_{\g k}-\mu & \sqrt{q_\sigma}V\\
    \sqrt{q_\sigma}V &  \epsilon_d  +q_\sigma t_d\xi_{\g k}-\mu+\lambda_\sigma
  \end{pmatrix}\Psi^\dagga
\end{equation}
where in above $\Psi^\dagger=\{ \hat c_{\g k,\sigma}^\dagger,\hat d_{\g k,\sigma}^\dagger\}$. Moreover $q_\sigma \big(\langle\hat n^d_\uparrow\hat n^d_\downarrow\rangle,\langle\hat n^d_\uparrow\rangle,\langle\hat n^d_\downarrow\rangle\big)$ is the usual Gutzwiller narrowing  factor \cite{Wysokinski2014,Wysokinski2014R} and 
\begin{equation}
\lambda_\sigma= \frac{\partial \langle \mathcal{H}\rangle_G}{\partial \langle n_{\sigma}\rangle_0}
\end{equation}
is a spin-resolved shift of the  chemical potential, both obtained in a self-consistent manner \cite{notka}. Here $\langle ...\rangle_0$ denotes an expectation value with the Slater determinant and $\langle ...\rangle_G$ with the Gutzwiller wave function under the Gutzwiller approximation.

In turn renormalized mean-field Hamiltonian for the AFM phase   defined in the reduced  Brillouin zone (RBZ) reads  
\begin{equation}
H_{AFM} =\sum_{\sigma,\g k \in RBZ}\!\! d\g k\ 
\Psi^\dagger
 \begin{pmatrix}
  \epsilon_{\g k}-\mu & 0  &\frac{q_{\sigma}+q_{\bar\sigma}}{2}V & \frac{q_{\sigma}-q_{\bar\sigma}}{2}V\\
 0 & \epsilon_{\g k+\g Q}-\mu & \frac{q_{\sigma}-q_{\bar\sigma}}{2}V&\frac{q_{\sigma}+q_{\bar\sigma}}{2}V\\
  \frac{q_{\sigma}+q_{\bar\sigma}}{2}V & \frac{q_{\sigma}-q_{\bar\sigma}}{2}V & \epsilon_d  +q_\sigma t_d\xi_{\g k}-\mu+ \frac{\lambda_{\sigma}+\lambda_{\bar\sigma}}{2}  & \frac{\lambda_{\sigma}-\lambda_{\bar\sigma}}{2} \\
  \frac{q_{\sigma}-q_{\bar\sigma}}{2}V&\frac{q_{\sigma}+q_{\bar\sigma}}{2}V & \frac{\lambda_{\sigma}-\lambda_{\bar\sigma}}{2}  & \epsilon_d  +q_\sigma t_d\xi_{\g k+\g Qs}-\mu +\frac{\lambda_{\sigma}+\lambda_{\bar\sigma}}{2} \\
 \end{pmatrix} \Psi,
\end{equation} 
where here $\Psi^\dagger=(c^\dagger_{\g k \sigma},c^\dagger_{\g k+\g Q \sigma},f^\dagger_{\g k \sigma},f^\dagger_{\g k+\g Q \sigma})$, and $q_\sigma$ and $\lambda_\sigma$ are obtained for a one sublattice.

Used variational method is well suited to capture a Fermi-liquid state, even though it disregards momentum-dependence of the quasiparticle weight. One can associate the approximation imposed on the solution of the $\mathcal{H}$ with a one concerning the $d$-electron self-energy
\begin{equation}
 \Sigma_\sigma(\omega,\g k)\simeq \Re\Sigma_\sigma(0)+\omega\frac{\partial\Re \Sigma_\sigma(\omega)}{\partial \omega}\Big|_{\omega\rightarrow0}.\label{self}
\end{equation}
Then the Gutzwiller factor $q_\sigma$ can be shown to play a role of a quasiparticle weight, ${z_\sigma\!=\!\big[1\!-\!\frac{\partial \Re\Sigma_\sigma(\omega)}{\partial \omega} \big|_{\omega\rightarrow0}\big]^{-1}}$.

We note that in principle a $p$-$d$ lattice model with the total-filling close to integer and in the large-$U$ limit can support  Mott or Kondo physics \cite{Coleman1984,Pruschke2000,Kotliar2008,Amaricci2008,Amaricci2009,Dong2013} being responsible for magnetism of localized moments.  However, for all discussed result  we have assumed total filling away from integer value, $n_t=1.6$, providing that for small-to-moderate interaction values the exchange interactions, and thus magnetism of localized moments is not entering present considerations. For that reason it is justified to use  Gutzwiller approximation that by the construction disregards exchange interactions \cite{Wysokinski2017R}.


\begin{thebibliography}{10}
\bibitem{notkas}
\bibinfo{note}{\mbox{Obtained} triple point is also supported at interaction -
  magnetic field plane due to the fact that Zeeman spliting strongly humpers
  AFM giving rise to FM2/FM1/AFM phase sequence.}
\end{thebibliography}

\section*{Acknowledgements}

Author thanks T. Dietl and T. Hyart for critical reading of the manuscript  as well as  C. Autieri, W. Brzezicki, P. Iwanowski and V. Taufour for stimulating discussions and extremely useful comments. 
Author also thanks prof. A. Green for  suggesting mentioned in the main text possible experimental indication that FM/AFM transition in \lcg\  takes place either due to proposed here mechanism or due to this relying on quantum fluctuations. 
The financial support from the Foundation for Polish Science under the ``START'' program is greatly acknowledged.   The  International Centre for Interfacing Magnetism and Superconductivity with
Topological  Matter  project  is  carried  out  within  the  International  Research  Agendas  program  of  the  Foundation  for Polish Science co-financed by the European Union under the European Regional Development Fund.  Author acknowledges the access to the computing facilities of the Interdisciplinary Center of Modeling at the University of Warsaw, Grant
No. G73-23 and G75-10.

\section*{Author contributions statement}

MMW did all the work. 

\section*{Additional information}

The author  declare no competing interests.

\end{document}